\documentclass[pre,aps,twocolumn,groupedaddress,amsmath]{revtex4}
\usepackage{graphicx}
\usepackage{multirow}
\usepackage{hyperref}

\begin{document}

\title{Defects in nematic membranes can buckle into pseudospheres}
\author{John R. Frank}
\email{jrf@mit.edu}
\affiliation{Massachusetts Institute of Technology}
\author{Mehran Kardar}
\affiliation{Massachusetts Institute of Technology}
\date{\today}

\begin{abstract}
A nematic membrane is a sheet with embedded orientational order, which can occur in biological cells, liquid crystal films, manufactured materials, and other soft matter systems.  By formulating the free energy of nematic films using tensor contractions from differential geometry, we elucidate the elastic terms allowed by symmetry, and indicate differences from hexatic membranes.  We find that topological defects in the orientation field can cause the membrane to buckle over a size set by the competition between surface tension and in-plane elasticity.  In the absence of bending rigidity the resulting shape is universal, known as a parabolic pseudosphere or a revolved tractrix.  Bending costs oppose such buckling and modify the shape in a predictable manner.  In particular, the anisotropic rigidities of nematic membranes lead to different shapes for aster and vortex defects, in principle enabling measurement of couplings specific to nematic membranes.
\end{abstract}
\maketitle

\section{Introduction}
By ``nematic membrane'' we refer to any flexible sheet incorporating ordered rod-like constituents.  For example, thin films of smectic-C liquid crystals are nematic membranes \cite{orsay1971, young78, spector93, shalaginov1996}.  Also, recently developed sheets of carbon nanotubes have nematic character \cite{islam2004, yu2007}.  Nematic order arises in lipid membranes with inclusions \cite{fournier1998a} and in the cell cytoskeleton, e.g. during mitosis \cite{fukui1991}.   Interestingly, \emph{in vitro} experiments on mixtures of cytoskeletal filaments and protein motors observe topological defects (asters and vortices), which spontaneously self-organize into a variety of patterns \cite{nedelec1997, surrey2001, lagomarsino2004}.  These experiments, and related simulations, use \emph{flat} geometries with various boundary conditions \cite{lee01, zumdieck05}.  Similar topological defects influence the shapes of real cells.  For example, cells of the alga Bryopsis sprout branches out of vortex-shaped defects that appear naturally in their cell wall of cellulose microfibrils \cite{green1960}.  To take a step toward understanding such living and \emph{in vitro} systems, we consider equilibrium shapes around defects in deformable nematic membranes.

We show that topological defects can buckle the membrane.  This has similarities to two other systems.  One is bulk nematic liquid crystals, which buckle into the third dimension around defect lines \cite{cladis72,meyer72,anisimov73} in a manner directly analogous to the shapes we find.  A second example is provided by deformable triangular latices, which have been studied extensively in the theory of two-dimensional melting.  While the physical picture is different, the model energy is equivalent to a nematic membrane with isotropic elastic constants.  Disclination defects culminating in a site with five or seven bonds (instead of the usual six) can lower their energy by buckling \cite{nelson87,seung88,nelson1994,nelson2002}.  When draped over curved surfaces, collections of such defects arrange in specific patterns \cite{vitelli04a,vitelli04b,vitelli06a, vitelli06b}.   If surface tension is neglected, five-fold disclinations assume an approximately cone-shaped form \cite{seung88}.

In contrast to the above cases, competition between the cost of surface area and rod misalignment determines the shape of the defects we consider.  When bending rigidity is neglected, we find that topological defects deform membranes into a simple universal shape known as a parabolic pseudosphere \cite{pseudosphere, kreyszig1991}.  The size (height and extent) of this universal form is governed by the ratio of surface tension to in-plane elasticity.  The inclusion of bending rigidity opposes this puckering.  If the bending cost is small, the singularities at the tip and rim of the defect become smoother.  The logarithmically diverging tip of the parabolic pseudosphere is replaced by the finite height of an \emph{elliptic} pseudosphere \cite{kreyszig1991}, and the sharp rim is replaced by an exponential falloff with a length scale related to rigidity.  Higher bending costs completely eliminate the buckling instability.  The anisotropic elasticity of nematics singles out specific defect orientations (asters and vortices); and corresponding anisotropies in bending rigidity lead to different length scales for their shapes.

The rest of the manuscript is organized as follows: In Sec.~\ref{sec:FreeEnergy}, we describe the free energy of a nematic membrane using tensor contractions from differential geometry.  This provides a compact formulation applicable to all deformations, including highly curved shapes.  In Sec.~\ref{sec:Defects}, we describe vortex and aster defects, derive shape equations for radially symmetric configurations, and solve them to find the buckled defect shapes.  Section \ref{sec:Conclusion} provides a summary and indications for future research.  In Appendix \ref{app:FixedGeometries}, we study filament orientations in fixed geometries, which may provide other ways of measuring the nematic membrane parameters.  In Appendix~\ref{app:Stability}, we check the linear stability of the buckled defect shapes.

\section{Elastic Free Energy of Nematic Membranes\label{sec:FreeEnergy}}
Using differential geometry to describe a two-surface in three-space, we construct a power series expansion for the free energy by selecting a linearly independent set of scalar contractions of the surface tensors.  For a surface described by an embedding vector ${\vec{X}(\sigma^1, \sigma^2)}$, one constructs tangent vectors, ${\vec{t}_i = \partial_i \vec{X}}$, by taking derivatives of the embedding vector with respect to its two parameters.  The metric tensor is then ${g_{ij} = \vec{t}_i \cdot \vec{t}_j}$.  The covariant derivative is defined such that ${D_i g_{jk}=0}$.  The curvature tensor is constructed from covariant derivatives of the tangent vectors as ${K_{ij} = (D_i \vec{t}_j) \cdot \hat{N}}$, where $\hat{N}$ is a surface normal.  One must choose a side to define the sign of $\hat{N}$.  In the principle directions basis, ${K_{ab} = \left( \begin{array}{cc} 1/R_1  & 0 \\  0  & 1/R_2 \end{array}  \right)}$, where $R_i$ are the radii of curvature \cite{david03}.

A unit-magnitude tangent vector field ${\vec{T}=T^i \vec{t}_i}$ represents the nematic particles.  At constant filament density, the magnitude ${T^i T_i = 1}$ is fixed and only its orientation changes \footnote{Relaxing ${T^iT_i=1}$ would introduce independent $TKKT$ and $T_\perp KKT_\perp$ terms and several new gradient terms.}.  Nematic symmetry implies invariance under ${T^i \rightarrow - T^i}$ \footnote{Dropping the nematic symmetry requirement introduces four new spontaneous curvatures, two of which are chiral.}.  A complete set of scalars up to second order in derivatives is
\begin{eqnarray}
{\cal F}_{nematic} & = & \sigma +  \frac{K_1}{2}  \left(D_i T^i\right)^2   + \frac{K_3}{2}   \left(D_i T^j_\perp \right)^2  \nonumber  \\
                             & + &  \frac{\kappa_{||}}{2} \left( T^i K_{ij} T^j  - H_{||}\right)^2  \nonumber \\
                             & + &  \frac{\kappa_{\perp}}{2}   \left( T^i_{\perp} K_{ij} T^j_{\perp} -H_{\perp} \right)^2  \nonumber \\
                             & + &  \frac{\kappa_{\times}}{2}   \left(T^i   K_{ij}   T^j_{\perp}  - H_{\times} \right)^2  \text{   .}  \label{eqn:NM}
\end{eqnarray}
This free energy density must be integrated with a surface area element ${dA = \sqrt{g}\,\,d^2\sigma}$, where $g$ is the determinant of the metric.   The weighted antisymmetric tensor ${\gamma_{ij} = \sqrt{g}\,\, \epsilon_{ij}}$ rotates one-tensors by $\pi/2$, such that ${T^i_{\perp} = \gamma^{ij} T_j}$  \cite{david03}.  Each term is manifestly positive, so stability demands that the moduli be positive.  In the remainder, we consider reflection symmetric, non-chiral membranes without spontaneous curvatures ${H_{||, \perp,\times}}$.  

Unlike parameterizations used to study nematic membranes near the hexatic fixed line \cite{nelsonpowers1992, powers1995}, this set of scalars cleanly delineates the anisotropic bending energies that make nematic membranes unique.  Creating more surface area costs $\sigma$ \cite{brochard1976,david91}.   In-plane splay and bend cost $K_1$ and $K_3$, which are the two-dimensional analogs of the bulk nematic Frank constants \cite{frank58,shalaginov1996}.   Membrane curvature in the direction of the local filament orientation costs $\kappa_{||}$.  Curvature perpendicular to the filaments costs $\kappa_{\perp}$.  These out-of-plane bending terms are the anisotropic analogs of the the Canham\cite{canham70}-Helfrich\cite{helfrich73} bending rigidity.  Saddle curves cannot be constructed from the other two out-of-plane bending terms and incur an independent energy cost of $\kappa_\times$.  The square of the chiral scalar $TKT_\perp$ \cite{nelsonpowers1992,nelson1993} is non-chiral.  The underlying membrane has a fluid character in that the particles can rearrange in the surface without stretching or shearing costs.

Compared to the \emph{splay}, \emph{bend}, and \emph{twist} of bulk nematics, nematic membranes have additional freedom that comes from relaxing a constraint: instead of three fields constrained to a unit vector, the nematic membrane constrains only two fields to a unit vector and allows a third field to range freely in describing the membrane's local deviation from flatness \footnote{Such constraints deserve further study in the spirit of Capovilla's and Guven's study of membranes with isotropic rigidity in Ref.~\cite{capovilla2005}.}.

In a system of motor proteins pulling on cytoskeletal filaments, $K_1$ would be proportional to motor density, which we assume to be uniform, and $\kappa_\perp$ would be determined primarily by the bare membrane's isotropic rigidity.  Filament rigidity would influence both $K_3$ and $\kappa_{||}$.  See Appendix \ref{app:FixedGeometries} for comments on $\kappa_\times$.

Perturbative RG near the hexatic fixed line \cite{powers1995} shows that thermal fluctuations reduce weak anisotropy, i.e., the three quantities ${\kappa_{\times}-\kappa_{||}-\kappa_\perp}$, ${\kappa_{||} - \kappa_\perp}$, and ${K_1 - K_3}$ fade at long distances, so that only the hexatic membrane energy remains, and
\begin{eqnarray}
{\cal F}_{hexatic} = \sigma + \frac{K_A}{2} (D_iT_j)(D^i T^j) + \frac{\kappa}{2}  \left( K^i_i - H_0\right)^2  \text{   ,}\label{eqn:hexatic}
\end{eqnarray}
where ${K_A = \frac{1}{2} (K_1 + K_3)}$ and ${\kappa=\frac{1}{2}(\kappa_{||}+\kappa_{\perp})}$.  
Under further rescaling, ${\kappa \rightarrow 0}$ and $K_A$ is unrenormalized.  Note that while the hexatic energy takes its name from the six-fold symmetry of triangular lattices, any n-atic symmetry with ${n \ge 3}$ restricts ${K_1 = K_3}$ and ${\kappa_\perp=\kappa_{||} = 2 \kappa_\times}$.   For polar ($n=1$) or nematic ($n=2$) membranes, the isotropic approximation is an important limiting case at one extreme of a phase diagram that deserves further attention.

Estimates of the thermal persistence length, $\xi_T$, of weakly anisotropic rigid membranes indicate an exponential form ${\log\xi_T \propto \kappa}$ \cite{powers1995, kleinert1986, gutjahr2006}.  Modest changes in $\kappa$ can thus sweep the persistence length from small values up to thousands of times the short-distance cut-off \cite{degennes1982}.  Effects unique to the nematic membrane can then appear in patches of material smaller than this persistence length.

\section{Buckled Defects Shapes\label{sec:Defects}}
\begin{figure}[ht]%
\includegraphics[width=125\unitlength]{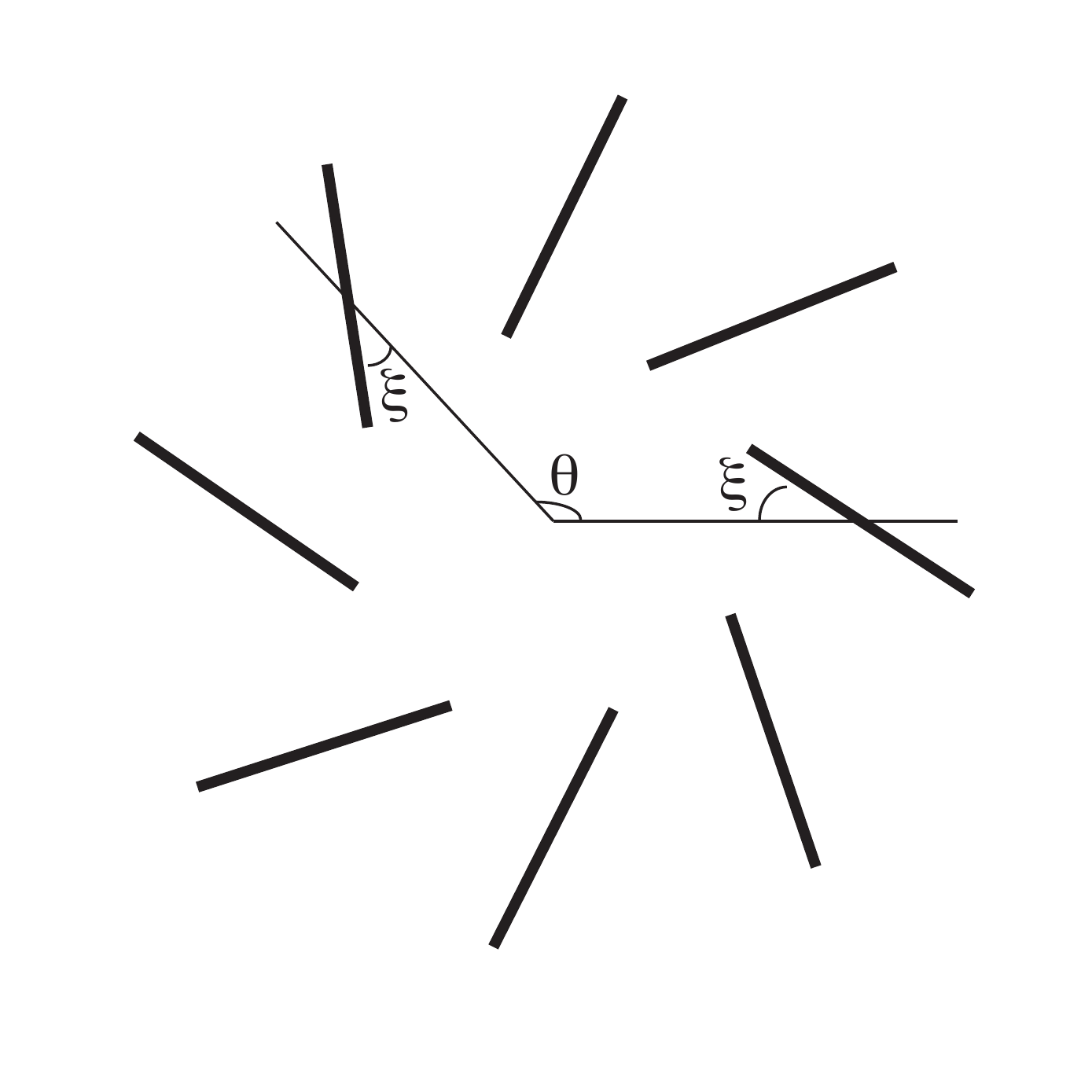}%
\caption{\label{fig:defectDiagram}Rod orientations around a general uniform +1 topological defect.  $\xi=0$ corresponds to an aster, and ${\xi=\pi/2}$ to a vortex.}%
\end{figure}

In the nematic phase, the rod orientation varies slowly throughout most of the material.  However, at particular defect points, the orientation may be undefined, because rods at neighboring locations point in all directions.  The \emph{topological charge} of a defect is the number of times that the orientation rotates through ${2 \pi}$ as the coordinate angle $\theta$ sweeps through ${2 \pi}$.  Different patterns appear for integer, half integer, and positive and negative charges.  The defect depicted in Fig.~\ref{fig:defectDiagram} is radially symmetric, and is rotated by a uniform angle $\xi$ with respect to the radial vectors.  

In the limit of isotropic rigidity, +1 defects with any radially uniform $\xi$ have the same energy.  The symmetry is removed by the anisotropic moduli in a nematic membrane, which distinguish asters ($\xi=0$) and vortices ($\xi=\pi/2$).  The energy of such a planar defect as a function of $\xi$ is
\begin{eqnarray}\label{defect-core}
E_{planar}(\xi) =\pi \left(K_1 \cos^2(\xi) + K_3 \sin^2(\xi)\right) \ln \frac{R}{a} + E_c(\xi) \text{,} 
\end{eqnarray}
where $R$ is the size of the membrane, $a$ is a short distance cutoff, and $E_c$ is a core energy reflecting the defect's microscopic situation inside of the core radius $a$. 

For ${K_3>K_1}$, asters have lower energy than vortices and are stable against in-plane deformations.  If ${K_1>2 \kappa_\perp}$, the defect energy is further reduced by buckling out of flatness to align the filaments in the third dimension.  (See Appendix~\ref{app:Stability} for linear stability analysis.)  Buckling comes at the expense of creating more area, so surface tension sets the size of the deformation.  Analogously, when ${K_1>K_3>2\kappa_{||}}$, vortices are stable and can reduce their core energy by tilting the surface around the defect.

To study this buckling, we minimize the nematic membrane energy, Eq.~(\ref{eqn:NM}), around fixed aster and vortex arrangements.  For a radially symmetric surface with no overhangs, we use the \emph{polar Monge tangent} representation with embedding vector ${\vec{X}(r, \theta)}$.  The height above the Monge plane is found by integrating the tangent angle $\chi(r)$ from a base value, so that
\begin{eqnarray}
\vec{X}(r,\theta) &=& \left( \begin{array}{l}
                                     r  \,\,  \cos(\theta)  \\
                                     r  \,\,  \sin(\theta)  \\
                                     \int^r{ \tan(\chi(r'))  dr'  }
                              \end{array} \right)  \text{   .} \label{eqn:polarmongetangent}
\end{eqnarray}
This yields a metric with no derivatives and thus lower order shape equations.  To handle shapes with overhangs, such as prolate vesicles \cite{seifert91}, one can parameterize the shape by contour length instead of Monge radius.  

The unit vector constraint is enforced by defining the angle $\xi$ such that
\begin{eqnarray}
T^i = \left( \begin{array}{c}
                        \cos(\xi) \cos(\chi) \\
                        \sin(\xi) / r
                    \end{array}
          \right)  \text{   .}  \label{eqn:unitOnPolarMongeTangent}
\end{eqnarray}
With this parameterization, the nematic membrane free energy becomes
\begin{widetext}
\begin{eqnarray}
F_{nematic}=2 \pi \int{ dr  \left(
\sigma \frac{r}{\cos(\chi)} + 
\left[
\begin{array}{l}
\kappa_{||} (\sin^2(\xi) \tan^2(\chi) + r^2 \cos^2(\xi) \chi'^2)  + \\
\kappa_{\perp} (\cos^2(\xi) \tan^2(\chi) + r^2 \sin^2(\xi) \chi'^2) + \\
\bar{\kappa}_{\times} \sin^2(\xi) \cos^2(\xi) ( \tan(\chi) - r  \chi')^2 +  \\
K_1  (\cos(\xi)  - r \sin(\xi) \xi')^2  
+ K_3  (\sin(\xi)  + r \cos(\xi) \xi')^2
\end{array} \right]  \frac{\cos(\chi)}{2 r}  \right) }  \text{   ,}
\end{eqnarray}
where ${\bar{\kappa}_{\times} = \kappa_{\times} - \kappa_{||} - \kappa_{\perp}}$.  We could have written the energy directly in terms of ${\bar{\kappa}_\times}$ by switching from a $(TKT)^2$ to a $TKKT$ parameterization as permitted by the unit-vector constraint.  Fixed aster or vortex configurations carry no energy cost from the term proportional to ${\bar{\kappa}_\times}$.  Setting to zero the functional derivative of $F_{nematic}$ with respect to $\chi$ yields a \emph{shape equation}, which for an aster (${\xi=0}$) becomes
\begin{eqnarray}
0=
\sigma \frac{r \sin(\chi)}{\cos^2(\chi)} 
+ \frac{\kappa_{||} }{2} (-2 \cos(\chi) \chi' + r \sin(\chi) \chi'^2 - 2 r \cos(\chi) \chi'') 
+ \frac{\kappa_{\perp}}{2} \left(1+\frac{1}{\cos^2(\chi)}\right) \frac{\sin(\chi)}{r}
- \frac{K_1}{2} \frac{\sin(\chi)}{r} \text{   .}  \label{eqn:shapeeqn}
\end{eqnarray}
\end{widetext}
For fixed vortices, the same shape equation holds after switching the coefficients ${\kappa_{\perp} \leftrightarrow \kappa_{||}}$ and ${K_1 \leftrightarrow K_3}$.  

For any membrane (hexatic or nematic) without stiffness ($\{\kappa\}=0$), defects have a simple universal shape resulting from the competition between the in-plane misalignment cost and surface tension.  The misaligned rods near the defect core can align by bending out of the plane into the third dimension, at the cost of increasing surface area.  The optimum tangent angle is given by the simple formula
\begin{eqnarray}
\cos(\chi) &=& \sqrt{\frac{2 \sigma r^2 }{K_1}}  = \frac{r}{r_0}  \text{   ,}  \label{eqn:hexaticsolution} 
\end{eqnarray}
where ${r_0 = \sqrt{K_1 / 2 \sigma}}$ is the distance outside of which surface tension dominates and flattens the surface.  Integrating the angle gives the universal shape
\begin{eqnarray}
h(r) &=&r_0 \left( \text{sech}^{-1}\left(\frac{r}{r_0}\right) - \sqrt{1-\left(\frac{r}{r_0}\right)^2} \right) \text{   ,} \label{eqn:hexaticsolutionHeight}
\end{eqnarray}
which approaches vertical at $r=0$ where the height is logarithmically divergent.  This may be regulated by a cut-off, such as the membrane thickness.  As a reference, at half the rim radius: ${h(r_0/2) \approx 0.45 \, r_0}$.  In a hexatic membrane ${K_1=K_3}$, so asters and vortices have the same radius.  In a nematic membrane, asters and vortices have different radii; the lower energy defect also has smaller size.   

This shape, Eq.~(\ref{eqn:hexaticsolutionHeight}), is known as a \emph{parabolic pseudosphere} or \emph{antisphere}, because it has constant negative Gaussian curvature equal to $-1/r_0^2$ \cite{pseudosphere, kreyszig1991}.  It is also known as a \emph{tractrisoid}, because it is half the surface of revolution generated by revolving a \emph{tractrix} about its asymptote \cite{tractrix}.  The tractrix is the path of an object being dragged by a string of constant length along a straight line that does not intersect the object.  Leibniz likened this problem to a dog owner dragging his hound by its leash and named the solution \emph{hundskurve}.  The hundskurve has been studied by Huygens and others \cite{lockwood1961}.  This construction makes it clear that the distance to the axis along the line tangent to any point on the surface is constant, i.e. the leash length is $r_0$.  These shapes of constant negative curvature are also known in quantum gravity as solutions to classical Liouville theory \cite{seiberg1990}.

This simple shape has singularities at the origin and at the rim ${r=r_0}$, which are modified by the membrane bending rigidities, $\kappa_\perp$ and $\kappa_{||}$, respectively.  Setting ${\kappa_{||}=0}$ removes all derivatives of $\chi$ from the shape equation, so a simple rearrangement provides the solution,
\begin{eqnarray}
\cos(\chi) = \sqrt{\frac{2 \sigma r^2 + \kappa_{\perp}}{K_1 - \kappa_{\perp}}} = \sqrt{\frac{\left(\frac{r}{r_1}\right)^2 + c}{1+c}} \text{   ,} \label{eqn:solution} 
\end{eqnarray}
where ${r_1 =  \sqrt{({K_1 - 2 \kappa_{\perp}})/{2 \sigma}}}$ is the new rim radius, and ${c={\kappa_{\perp}}/({K_1-2\kappa_{\perp}})}$ is related to the now finite slope at the tip.  For sufficiently large $K_1$, the surface puckers out of the plane for ${r<r_1}$, with a profile
\begin{eqnarray}
h(r) &=& \int^{r_1}_r{\tan\left[\pm \cos^{-1}\left( \sqrt{ \frac{2 \sigma r^2 + \kappa_{\perp}}{K_1 -  \kappa_{\perp}}}\right) \right] dr}  \\
      &=&\pm  r_1  \int_{ \frac{r}{r_1}}^1 { \sqrt{\frac{1-u^2}{c + u^2}} du} \text{   .}  \label{eqn:ellipticIntegral}
\end{eqnarray}
For ${K_1< 2 \kappa_\perp}$ or for ${r_1<r}$, this solution is not real, so $\chi=0$ becomes the only solution to the shape equation.  

\begin{figure}[ht]%
\includegraphics[width=250\unitlength]{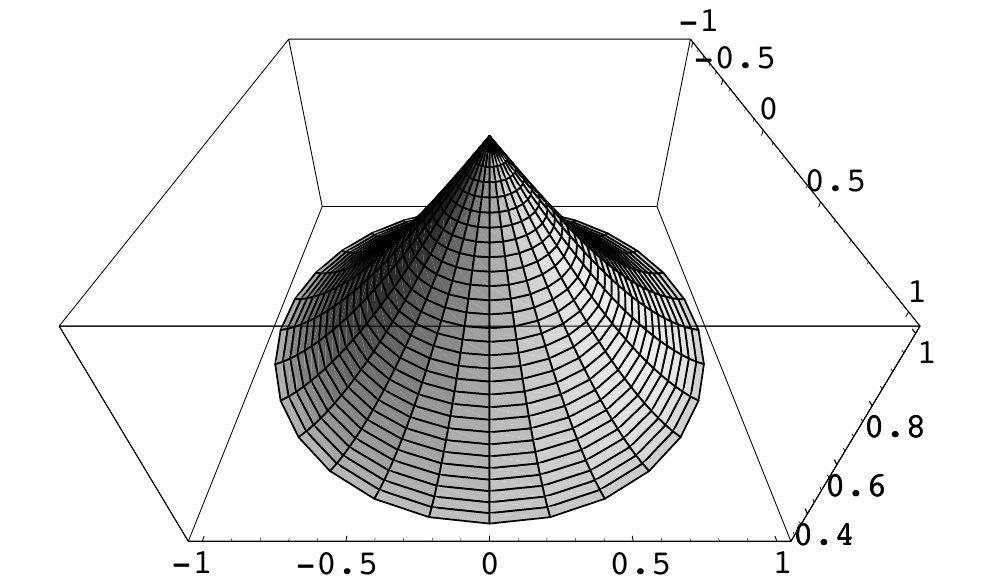}%
\caption{\label{fig:peak}An example of buckled shape for $r_1=1$ and $c=1$.}%
\end{figure}

Equation~(\ref{eqn:ellipticIntegral}) is a complete elliptic integral of the second kind \cite{EllipticE}.  We change variables $u \rightarrow \sqrt{c} \sinh(u)$ to obtain
\begin{eqnarray}      
h(r) &=& \pm  r_1  \sqrt{1+c} \int{ \sqrt{1 - \frac{c  \cosh^2(u)}{1+c}} \, du }    \text{   ,} \label{eqn:ellipticPseudosphere}
\end{eqnarray}
where the integration ranges from $\sinh^{-1}(r/(r_1\sqrt{c}))$ to $\sinh^{-1}(1/\sqrt{c})$.  In the study of surfaces with constant Gaussian curvature, Eq.~(\ref{eqn:ellipticPseudosphere}) is a familiar expression for an \emph{elliptic} pseudosphere  \cite{kreyszig1991}.   Figure~\ref{fig:peak} shows an example elliptic pseudosphere.  

The bending rigidity cuts off the logarithmically diverging tip near the core.  Near the origin, the elliptic pseudosphere is approximately cone-shaped with slope ${\sqrt{ 1 / c}=\sqrt{(K_1-2\kappa_\perp) / \kappa_\perp }}$,
\begin{eqnarray}
h(r) \rightarrow \pm  \frac{r}{\sqrt{c}}  \text{   .} \label{eqn:cone} 
\end{eqnarray} 
If the $r^2$ term in Eq.~(\ref{eqn:solution}) were not present, the shape would be a cone.  Unlike the cone, pseudospheres have constant Gaussian curvature,
\begin{eqnarray}
\frac{1}{R_1 R_2} &=& \det \left[K_i^j\right] = \det \left[ \hat{N} \cdot  \partial_i \vec{t}^j  \right]  \\
                             &=& - \frac{1}{r_1^2 (1+c)}  \text{   ,}
\end{eqnarray}
where $\hat{N}$ is the unit vector proportional to $\vec{t}_1 \times \vec{t}_2$ and we have carried through the computation after inserting the coordinate tangent vectors for the elliptic pseudoshere,
\begin{eqnarray}
\vec{t}_1 &=& \left(  \cos(\theta) ,  \sin(\theta) ,  \sqrt{\frac{1-\left(\frac{r}{r_1}\right)^2}{c+\left(\frac{r}{r_1}\right)^2}}   \right) \\
\vec{t}_2 &=& \left(  -r \sin(\theta) , r \cos(\theta) , 0  \right) \text{  .}
\end{eqnarray}

Even with finite $\kappa_\perp$ the above shape retains a cusp-like singularity at the origin.  We may well question how the singularity is modified by inclusion of cut-offs and higher order terms.  A simple short-distance cut-off, $a$, can be introduced as the radius of a hemispherical or similar cap over the singular point at the origin.  The curvature energy density ${\sim 1/a^2}$ integrated over the cap's area ${\sim a^2}$ leads to a finite energy.  We can then regard this as a benign singularity that adds a constant to the defect core energy $E_c$ in Eq.~(\ref{defect-core}).

Substituting Eq.~(\ref{eqn:solution}) into the full shape equation, Eq.~(\ref{eqn:shapeeqn}), leaves a term proportional to both $\kappa_{||}$ and to $r$, so the elliptic pseudosphere is expected to remain valid as ${r\to0}$ near the core.  The situation at the rim is very different: Designating the distance from $r_1$ by ${\epsilon = 1 - {r}/{r_1}}$, one sees that ${\chi \propto \sqrt{\epsilon}}$ as ${\epsilon \to 0^+}$ and is zero immediately outside this radius.  The abrupt rim would cause the energy proportional to $\kappa_{||}$ to diverge, so when ${\kappa_{||}>0}$, the defect shape must be different.  Since $\chi$ tends to zero away from the core, we linearize the shape equation for small $\chi$ and $\chi'$ to
\begin{eqnarray}
0 &\approx& \left((K_1-2 \kappa_\perp)\frac{1}{r} - 2  \sigma r\right) \chi + 2 \kappa_{||} (\chi' + r \chi'')  \text{  .}\label{eqn:chidot}    
\end{eqnarray}
After changing variables to $\epsilon$ and redefining ${\chi \rightarrow \chi\left( \epsilon \right)}$ to be a function of $\epsilon$, the linearized shape equation is
\begin{eqnarray}
0 &=&  \epsilon  (\epsilon - 2) \chi - \frac{ 2 \kappa_{||} (\epsilon-1) }{K_1 - 2 \kappa_{\perp}}   \left( \chi' + (\epsilon-1)  \chi'' \right)  \text{   .}
\end{eqnarray}
Note that the approximation is made for small $\chi$ and $\chi'$, and $\epsilon$ need not be small.  For real-valued $\chi(\epsilon)$, this equation is solved by modified Bessel functions of the second kind with imaginary order.  The order and argument both diverge with vanishing $\kappa_{||}$, as
\begin{eqnarray}
\chi(\epsilon) &\propto& K\left[ i \nu,  \nu \frac{r}{r_1} \right]  \text{   ,}  \label{eqn:longtailofeta} 
\end{eqnarray}
with
\begin{eqnarray}
\nu &=&  \sqrt{\frac{K_1 - 2 \kappa_{\perp}}{ 2 \kappa_{||}}} \text{   .}
\end{eqnarray}   
This solution decays exponentially and has no zeros for ${r_1 \le r}$.  Since our parameterization does not handle overhanging surfaces, $\chi$ is limited to the range ${(-\pi/2, \pi/2)}$.  Thus, for a given value of $\nu$, the amplitude must be such that the solution stays in this range.  For $\nu$ of order one and larger, an amplitude of unity yields a $\chi$ that is sufficiently small for ${r_1 \le r}$ that the linearized shape equation is valid.  It approaches zero asymptotically, so the rim radius at which ${\chi=0}$ shifts to infinity.  The asymptotic form of Eq.~(\ref{eqn:longtailofeta}) is \cite{balogh1967, abramowitz1972}
\begin{eqnarray}
\chi  \sim  \frac{ e^{-  \sqrt{\frac{3}{2}} \nu  \frac{r}{r_1}   }}  {\sqrt{\frac{r}{r_1}  \nu }}    \text{   ,   for  }  (\nu, r) \longrightarrow  \infty  \text{   ,}
\end{eqnarray}
which shows that bending rigidity introduces a new length scale
\begin{eqnarray}
r_2 \equiv\sqrt{\frac{2}{3}} \frac{ r_1}{ \nu} = \sqrt{ \frac{2 \kappa_{||} }{3 \sigma} }  \text{   .}
\end{eqnarray}

\begin{figure} [ht]%
\includegraphics[width=2.75in]{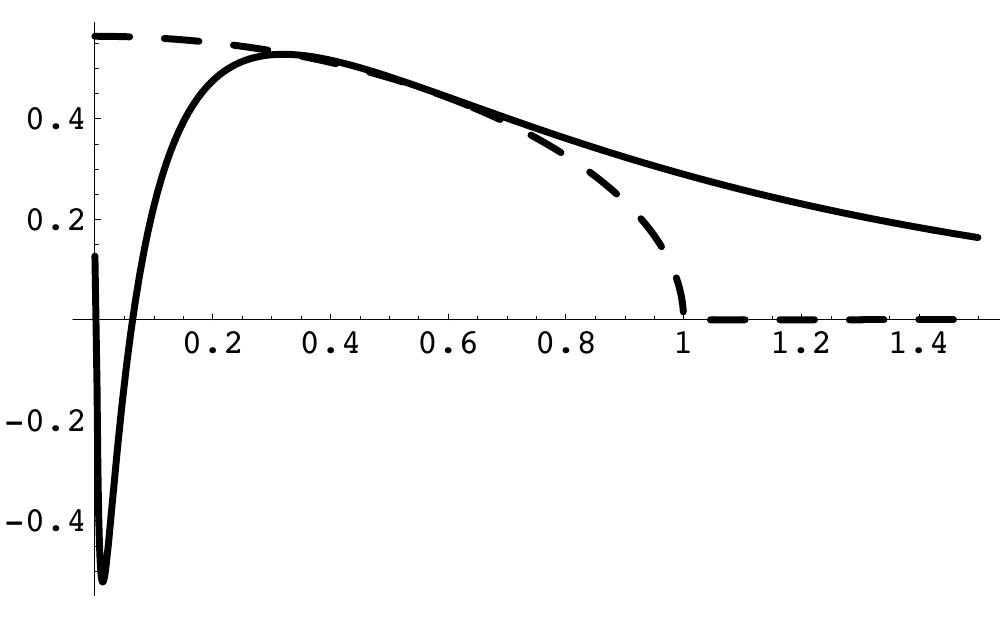}%
\caption{The tangent angle (not the height) as a function of radius in units of $r_1$.  Solid curve: The Bessel function solution for $\chi(\epsilon)$ for $\nu = 1$.  Dashed curve: $\cos^{-1}\left( \sqrt{{\left(\left(r/r_1\right)^2 + c\right)}/{(1+c)}}  \right)$ for $c= 2.5$ chosen to suggest matching in the crossover. \label{fig:near-transition}}%
\end{figure}%

As shown in Fig.~\ref{fig:near-transition}, this solution for $\chi$ oscillates sharply near the core, which invalidates the small $\chi'$ approximation.  In this region, nonlinearities take over and the shape crosses over to the elliptic pseudosphere.  As $\kappa_{||}$ approaches zero, the Bessel function becomes flat for $r_1<r$ and oscillates rapidly inside this radius, thus restoring the abrupt rim.

\section{Conclusions and Outlook\label{sec:Conclusion}}
We predict that topological defects can buckle nematic membranes into elliptic pseudospheres with exponentially decaying rims.  Let us explore this possibility in the case of a freely suspended thin film of smectic-C liquid crystal.  The observation that these films maintain orientational order at room temperature suggests that $K_1$ (and $K_3$) are at least of the order of ${k_B T_{\text{room}}\approx 4 \times 10^{-14} \text{erg}}$.  If we assume a surface tension typical of \emph{bulk} liquid crystal interfaces, $\sigma \sim 10 \text{ ergs}/\text{cm}^2$, and ignore bending rigidities ($\{\kappa\}=0$), then from $K \sim k_B T_{\text{room}}$, we would estimate a rim radius $r_0=\sqrt{K/2\sigma}\approx 1/2 \text{ nm}$, which is smaller than a typical film thickness and beyond the limits of this coarse grained model.  To create larger (observable) defects requires either smaller surface tension or larger in-plane stiffness than this initial estimate.  Some physical systems may allow this.

For example, studies of thin films of liquid crystal often observe a small surface tension, because the chemical potential for particles in the film is similar to the chemical potential in the meniscus surrounding the film \cite{brochard1976}.  This reservoir on the edge of the suspended film allows the film to increase its area at a low energetic cost.  One might control the size of buckled defects by manipulating the surface tension via this reservoir.  

In addition to having a surface tension smaller than our initial estimate, some materials have observed values of $\{K\}$ one or two orders of magnitude larger than room temperature.  For example, scattering studies by Spector et al on thin films of smectic-C 8OSI found large values of $\{K\}$ and surface tension made small by the meniscus \citep{spector93}.  This particular study used a smectic-C tilt angle of $32.2^\circ$ and found $K_1 / K_3 = 4.6(\pm0.4)$, $\kappa_{||}/K_3 = 3.4(\pm0.3)$, $\kappa_\times/\kappa_{||} = 5.5(\pm3.1)$, $\kappa_\perp/\kappa_{||}=75(\pm24)$, and $\kappa_{||} \approx 10^{-12} \text{ergs}$.  Unfortunately, the large value of $\kappa_{||}/K_3$ prevents buckling.  Since $K_1/K_3 > 1$, flat vortices should be stable relative to asters.  For other tilt angles or other materials, one might hope to find lower values of $\kappa_{||}/K_3$ that allow buckling.  

For sufficiently floppy films, buckled defects could be observable via specular reflection or by interferometry techniques used to measure the flatness of mirrors.  Additionally, islands of smectic-C material may provide means of manipulating single defects with laser tweezers, although coupling between the island's multiple smectic layers may introduce additional affects \cite{pattanaporkratana2003}.

While smectic-C 8OSI has sufficiently small surface tension and large in-plane stiffness, its bending rigidity suppresses buckling.  That such shapes have not been observed so far in other materials may well be an indication of the importance of bending rigidity.  Since typical lipid membranes have $\kappa\sim5k_BT_{\text{room}}$, this is a severe constraint.  We note, however, that for stiff rods (nanotubes, cytoskeletal filaments) embedded in membranes, the rigidities, $\kappa_\parallel$ and $\kappa_\perp$, and corresponding stiffnesses, $K_1$ and $K_3$, may well differ by orders of magnitude.  The challenge remains to obtain estimates of these parameters for specific microscopic models, and come up with an appropriate system for the study of buckled defects.  Observations of these shapes in nematic membranes may provide estimates of the ratios between $\kappa_\perp$, $\kappa_{||}$, $K_1$, $K_3$, and $\sigma$.   Measurements of $\kappa_\times$ are possible via other shapes as described in Appendix \ref{app:FixedGeometries}.

If one could control the anisotropic bending rigidities individually, one might be able to sweep a nematic membrane through a sequence of regimes in which different types of defects are stable.  For example, for ${K_3<K_1}$, if one could hold $\kappa_{||}$ fixed while adjusting $\kappa_\perp$ one might observe buckled vortices when ${2\kappa_\perp < K_1 - K_3 + 2 \kappa_{||}}$, and buckled asters when ${K_1-K_3+2 \kappa_{||}<2 \kappa_\perp < K_1}$, and buckled vortices again when ${K_1<2 \kappa_\perp}$.  By increasing $\kappa_{||}$ while keeping $\kappa_\perp$ in any of these regimes, one would flatten the preferred shape of vortices.  Thus, it is possible for vortices and asters to prefer buckled or flat shapes independently.

In focusing on shapes of minimal energy, we have neglected thermal fluctuations.  At long distances, thermal fluctuations reduce the differences between aster and vortex defects in \emph{weakly} anisotropic membranes \cite{powers1995}.  In future work, we would like to explore if this is still the case in \emph{strongly} anisotropic membranes, or if thermal fluctuations can enhance the anisotropy.

Unlike bulk nematics, nematic sheets often appear with naturally periodic boundaries such as closed vesicles.  By the Poincare-Brouwer theorem \cite{kamien2002}, a genus zero nematic vesicle must have topological charge of +2.  In fact, defects can burst the vesicle \cite{mackintosh91, park96, evans96}.  This resembles Bryopsis sprouting branches out of defects in its tethered nematic cell wall -- a topic to which we hope to return in the future.

\begin{acknowledgements} 
We thank Jacques Dumais, Vincenzo Vitelli, Noel Clark, Matthew Headrick, and Ari Turner for helpful conversations.  JRF thanks the Fannie and John Hertz Foundation for support.  This work was supported by the NSF through grant DMR-04-26677 (MK).
\end{acknowledgements}

\appendix
\section{Nematic Membranes in Fixed Geometries\label{app:FixedGeometries}}
While ${\kappa_\times}$ did not contribute to the shape of +1 defects, it affects other geometries.  Following de Gennes' molecular field argument \cite{deGennes93}, we impose the unit vector constraint via a Lagrange multiplier $\lambda$ and seek energy minimizing filament configurations in fixed geometries.  In a two-bein basis aligned with the principle directions, the curvature tensor is diagonal and ${T^i = e^i_a T_a}$, where $e^i_a$ is a transformation to local coordinates in which ${g_{ij} = \delta_{ij}}$ at each point \cite{david03}.  In the principle two-bein, 
\begin{eqnarray}
	T_a = \left(  \begin{array}{c}\cos(\xi) \\ \sin(\xi) \end{array} \right) \text{.}
\end{eqnarray}
	In the following, there is no summation over $a$ or ${\bar{a} = (a + 1) \mod (2)}$.   The functional derivative of Eq.~(\ref{eqn:NM}) in the principle two-bein reads
\begin{eqnarray}
	H_a  &\equiv& \frac{\delta F}{\delta T_a} = - \lambda(\sigma^1, \sigma^2) T_a \\
	        &=& T_a  \left\{ \kappa_{||}  C_a^2 + \kappa_{\perp} C_{\bar{a}}^2 + \bar{\kappa}_{\times} T_{\bar{a}}^2( C_1 - C_2)^2 \right\} \\ \nonumber
	            &&   - K_1 \partial_a  \left(   D_c T_c)  \right)   - K_3 \epsilon_{ba} \partial_b   \left(   \epsilon_{cd} D_c T_d  \right) \\
	        &=& T_a \left\{ \kappa_{||}  C_a^2 + \kappa_{\perp} C_{\bar{a}}^2 + \bar{\kappa}_{\times} T_{\bar{a}}^2( C_1 - C_2)^2 \right\} \\ \nonumber
	            &&   - K_1 \partial_a  \left( \partial_c\xi - A_c  \right) T_{\perp c}   \\ \nonumber
	            &&  - K_3 \epsilon_{ba} \partial_b  \left( \partial_c\xi - A_c  \right) T_{c}  \text{ ,}
\end{eqnarray}
	where $C_a$ are the principle curvatures and ${A_a=\hat{e}_1 \cdot \partial_a \hat{e}_2}$ is the spin connection.  The three-vectors ${\hat{e}_a = e^i_a \vec{t}_i}$ form an orthonormal basis in the principle two-bein.  To setup the molecular field equation, one must carry out the derivatives and pullout an overall factor of $T_a$ to obtain an expression for $\lambda$ that is a function of index $a$.  One obtains an equation for $\xi$ by requiring $\lambda$ to be a scalar, i.e., to have the same value for both $a=1$ and $a=2$.  Solutions to this equation for $\xi$ extremize the energy.  

Considering first ${K_1 = 0 = K_3}$, the equation yields a simple solution for $\xi$,
\begin{eqnarray}
\cos^2(\xi) = \frac{1}{2}\left(1 + \frac{\kappa_{||} - \kappa_\perp}{\kappa_\times - \kappa_{||} - \kappa_\perp} \frac{C_1 + C_2}{C_1-C_2} \right) \text{  .} \label{eqn:molecularFieldEqnNoK}
\end{eqnarray}
This is only valid with both components of $T_a$ are non-zero, so $\xi=0$ and $\xi=\pi/2$ must also be considered in the list of possible $\xi$ values.  One must check which candidate value for $\xi$ minimizes the energy for particular values of $\{\kappa\}$ and the principle curvatures.  In the following, we list a few special cases.  When ${\kappa \equiv \kappa_{||} = \kappa_\perp \ne \kappa_\times / 2}$ and ${C_1 \ne C_2}$, one has that  ${\xi= \pi / 4}$ minimizes the energy if
\begin{eqnarray}
\frac{ \kappa }{\kappa_{\times} } > - \frac{\left( C_1 - C_2 \right)^2 }{4 C_1 C_2}  \text{  .}
\end{eqnarray}
Otherwise, the filaments align with the least curved direction.

When ${\kappa_\times=\kappa_{||} + \kappa_\perp}$, so that ${\bar{\kappa}_\times=0}$, the orientation can be found by minimizing the energy with respect to $\xi$ directly, instead of the molecular field equation.  The result for $K_1=0=K_3$ and $C_1 < C_2$ is shown in Table~\ref{tab:stabcriteria}.

\begin{table}[h]
\caption{\label{tab:stabcriteria}}
\begin{tabular}{|rl|l|}
\hline
& Angle& Stability Criterion \\  \hline
$\xi$&$=0$ &          $C_1 \kappa_{||} > C_2 \kappa_\perp$ \\ \hline
$\xi$&$=\pi/2$ &     $C_1 \kappa_\perp > C_2  \kappa_{||}$ \\ \hline
\multirow{2}{*}{$\cos^2(\xi)$} &\multirow{2}{*}{$=\frac{C_1 \kappa_\perp - C_2  \kappa_{||}}{(C_1 - C_2)( \kappa_\perp + \kappa_{||} ) }$} & $C_1 \kappa_\perp < C_2 \kappa_{||}$,  \\
&&  $C_1 \kappa_{||} < C_2 \kappa_\perp$  \\  \hline
\end{tabular}  
\end{table}

On a developable surface,  i.e. ${C_1=0}$ and ${C_2\ne0}$, when ${\kappa_{\times} \ge 2  \min\left\{ \kappa_{||}, \kappa_{\perp} \right\}}$ the stable orientation is aligned with the uncurved direction.  For smaller values of ${\kappa_\times}$, a special intermediate angle is the global minimum,
\begin{eqnarray}
\cos^2(\xi) =  \frac{2 \kappa_{||} - \kappa_{\times} }{2 \left( \kappa_{||} + \kappa_{\perp} - \kappa_{\times} \right) } \text{  .}
\end{eqnarray}
Note that this only occurs when \emph{both} parallel \emph{and} perpendicular bending are more costly than ${\kappa_\times/2}$.  This might result from rods that weaken the sheet or have a specific texture on the rod's surface.

	For a developable surface, the spin connection is zero, so the covariant derivatives become regular partial derivatives.  Thus, on a cylinder, far from boundaries, a constant orientation solves the full molecular field equation with the gradient terms included.  This could allow experimental measurement of  ${( 2 \kappa_{||}  -\kappa_\times )/(\kappa_{||} + \kappa_{\perp} - \kappa_{\times} )}$.

	In more general geometries, in-plane splay and bend compete with out-of-plane bending in a non-linear PDE, which, in principle, can be numerically integrated to fit model parameters to vectorized images of a real nematic membrane.  Computing model parameters from such images in the presence of topological defects requires care.

\section{Stability of Buckled Defect Shapes\label{app:Stability}}
As discussed in Appendix~\ref{app:FixedGeometries}, the relative strength of $\kappa_\times$ plays an important role in the stability of orientation patterns on curved shapes.  Substituting the principle curvatures for the elliptic pseudosphere into Eq.~(\ref{eqn:molecularFieldEqnNoK}) yields an equation for $\xi$ that is not constant,
\begin{eqnarray}
\cos^2(\xi) = \frac{2 \kappa_\perp \left(\frac{r}{r_1}\right)^2 - \kappa_\times}{2(\kappa_\perp - \kappa_\times)} \text{  ,} 
\end{eqnarray}
and thus not the perfect aster (or vortex) that we assumed when setting up the shape equation, Eq.~(\ref{eqn:shapeeqn}).  Since Eq.~(\ref{eqn:molecularFieldEqnNoK}) was derived assuming $K_1=0=K_3$, the question remains whether the buckled defect is stable to perturbations away from a perfect aster (or vortex).

To check this, we construct linearized evolution equations for small perturbations, 
\begin{eqnarray}
\frac{d}{dt}\left(\begin{array}{c}\Delta \\ \Xi\ \end{array} \right) &\propto& -  \frac{\delta F}{\delta (\Delta, \Xi) } 
      \approx M \left(\begin{array}{c}\Delta \\ \Xi \end{array} \right) \text{,}
\end{eqnarray}
where $\Delta$ represents deviations of the surface away from a pseudosphere, and $\Xi$ represents deviations away from an aster ($\xi=0$).  Perturbations of the height field couple with perturbations of the angle field, so all four components of the two-by-two matrix of differential operators, $M$, are non-zero.  The perturbations are functions of both radius and angle, and are generally not radially symmetric.  To solve this, we write the perturbations in a Fourier basis, 
\begin{eqnarray}
\left( \begin{array}{c}\Delta(r,\theta,t) \\ \Xi(r,\theta,t)   \end{array} \right) &=&  \sum_m{ \left( \begin{array}{c} \Delta_m(r) \\ \Xi_m(r) \end{array} \right)  e^{i m \theta}   e^{\lambda_m t} } \text{  ,}
\end{eqnarray}
where each two-vector $(\Delta_m, \Xi_m)$ is independent.  Substituting this solution into the evolution equation gives a separate set of coupled equations for each $m$-value. 

Neglecting bending rigidity, and choosing units of energy such that $\sigma=1$ and units of length such that $K_1=2 \sigma$, we have for each value of $m$,
\begin{widetext}
\begin{eqnarray}
M^{(m)}= \left[ \begin{array}{c|c} \frac{r (-K_3 m^2 r + 2 (1-r^2)((2-4r^2) \partial_r +r (1-r^2) \partial_r^2) ) }{1-r^2} &
       \frac{i m ( (2-K_3) r  - (2-(2-K_3)r^2) \partial_r ) }{\sqrt{1-r^2}} \\ \hline 
      \frac{i m ( K_3 r + (1-r^2)(2-(2-K_3)r^2) \partial_r) }{r(1-r^2)^{3/2}} & - \frac{2 m^2}{r^4} + K_3 \partial_r^2
    \end{array} \right]  \text{.} \label{eqn:evoMatrix}
\end{eqnarray}
\end{widetext}

For $m=0$, the equations decouple.  Since these are perturbations, we must find real-valued solutions that vanish at the boundaries $r=0$ and $r=1$.  The equation for $\Xi$ has such a solution,
\begin{eqnarray}
\Xi \propto \sin\left( r \sqrt{ \frac{-\lambda_0}{K_3}}\right) \text{   ,}
\end{eqnarray}
if $\lambda_0 = - K_3 \pi^2 n^2$ for integer $n$.  This is always negative.  The equation for $\Delta$ has the real-valued solution
\begin{eqnarray}
\Delta& \propto &r^{2a}\, _2F_1(a,a+\frac{3}{2};2a+\frac{3}{2};r^2)  \nonumber\\
&& + r^{2b} \, _2F_1(b,b+\frac{3}{2};2b+\frac{3}{2};r^2)  \text{   ,}
\end{eqnarray}
where $\,_2F_1$ is the Gauss hypergeometric function and ${a=(-1-\sqrt{1+2 \lambda_0})/4}$ and ${b=(-1+\sqrt{1+2 \lambda_0})/4}$.  Since the third argument exactly equals the sum of the first two, $0<r<1$ is the convergent domain for these functions.  An ad hoc numerical study indicates that ${\lambda_0 \rightarrow - \infty}$ might extend this domain and allow the limit ${\Delta(1)\rightarrow0}$.  These functions also diverge at $r=0$, and again a large negative $\lambda_0$ appears to mediate this because the function oscillates rapidly and might average to zero as ${r\rightarrow0}$.  We lack an analytic treatment of this asymptotic regime, so we turn to a numerical method below. 

Considering $0<m$ and substituting $u=r^2$, one sees that Eq.~(\ref{eqn:evoMatrix}) consists of second-order ODEs with non-essential singularities at two points (the boundaries), so the equations can be transformed into hypergeometric differential equations \cite{hypergeomericdifeqn}.  By combining linearly independent solutions, one might construct real solutions that meet the boundary conditions for all values of $m$.  After satisfying these constraints, one would obtain expressions for $\lambda_m$, which, when negative, indicate stable regions of parameter space.  This approach is complicated even when rigidity is neglected.

Instead of taking this approach, we have checked stability numerically by discretizing the fields.  We represent the deviations of the height and angle fields by a large column vector of field values at discrete steps in radius and polar angle.  By representing the derivative operators as banded square matrices acting on this large vector, one obtains a matrix of numbers for any given set of parameter values.   The largest non-zero eigenvalue of this matrix determines the stability of the shape.  If the largest non-zero eigenvalue is negative, then that set of parameters suppresses perturbations and the shape remains stable. 

We have carried out such a numerical procedure.  Generally, the buckled aster is stable for $2 \kappa_\perp < K_1<K_3$ and any $0<\kappa_\times$.  The analogous statement holds for buckled vortices.

\bibliography{nematic-membranes}

\end{document}